\documentclass[aps,prl,showpacs,showkeys,twocolumn,amsmath,amssymb]{revtex4}

\usepackage{graphicx}
\usepackage{dcolumn}
\usepackage{natbib}
\usepackage{color,calc}
\usepackage{bm}

\newcommand{\bvs}{BaVS$_{3-\delta}$}
\newcommand{\bsvs}{Ba$_{1-x}$Sr$_x$VS$_3$}
 
\begin{document}

\title{Evidence of ferromagnetic quantum phase transition in \bsvs~}
\author{Andrea Gauzzi$^1$}
\email{andrea.gauzzi@upmc.fr}
\author{Neven Bari\v{s}i\'{c}$^2$}
\author{Francesca Licci$^1$}
\author{Gianluca Calestani$^3$}
\author{Fulvio Bolzoni$^1$}
\author{Patrik Fazekas$^4$}
\author{Edi Gilioli$^1$}
\author{L\'{a}szl\'{o} Forr\'{o}$^2$}

\affiliation{$^1$IMEM-CNR, Area delle Scienze, 43010 Parma, Italy\\
$^2$Institut de Physique de la Mati\`ere Complexe, EPFL, CH-1015 Lausanne, 
Switzerland\\
$^3$Dipartimento di Chimica, Universit\`a di Parma, Area delle Scienze, 43100 
Parma, Italy\\
$^4$Research Institute for Solid State Physics and Optics, P.O. Box 49, H-1525 
Budapest, Hungary}

\date{\today}

\begin{abstract}
Resistivity under pressure and magnetization measurements on \bsvs~ single 
crystals with 0 $\leq x \leq$ 0.18 and no sulphur deficiency show an abrupt 
onset of ferromagnetic (FM) order at a critical value $x$=0.07, concomitant to a 
change of the magnetic properties at the metal-insulator transition (MIT) and to 
a collapse of the unit cell at ambient temperature. A reduction of the MIT 
temperature to 50 K upon $x$ that scales as the V-S distance is also observed. 
This gives evidence of a chemical pressure induced quantum phase transition that 
stabilizes the incipient FM order of BaVS$_3$. The $0.07<x$ results suggest a 
coexistence of FM and metallic phases at larger $x$.
\end{abstract}


\maketitle

The hexagonal perovskite \bvs \cite{gar} displays a manifold of magnetic, 
transport and structural properties driven by 3$d$(V$^{4+}$) electron correlations. 
In the high temperature metallic phase, it consists of hexagonal closed 
packed chains of face-sharing VS$_6$ octahedra. The MI transition $T_{MI}$=69 K 
is preceded by a zigzag of the chains at 240 K leading to an orthorhombic 
structure \cite{tak}. $T_{MI}$ decreases under pressure and the insulating phase 
is suppressed for $P>2$ GPa \cite{for}. In stoichiometric compounds 
($\delta$=0), the ambient pressure transition is concomitant to an AF-like cusp 
of the magnetic susceptibility, $\chi$ \cite{gar}, and to a $2 \times c$ 
superstructure along the chains \cite{ina,fag}. A neutron scattering study 
\cite{nak} reported a long range AF-like order in the $ab$-plane and an intrachain 
ferromagnetic (FM) one at $T_X$=30 K, related to the cusp. Assuming a large number 
of degenerate configurations of singlet pairs with different orbital symmetry in 
the triangular plane, a picture of spin-orbital liquid was proposed \cite{nak,mih}. 
A latent FM instability is apparent from the positive Weiss constant in the metallic 
phase and the FM behavior of sulphur-deficient samples \cite{mas}. In agreement with 
these experiments, \emph{ab initio} calculations \cite{xue,san} predict a 
dominant FM coupling along $c$ and a comparatively weak AF one in the $ab$-plane, 
thus suggesting a ground state of FM chains with a nearly frustrated AF interchain 
coupling.

The origin of the MI transition and its relation with the magnetic and orbital 
orderings remain controversial. Overlapping wide and narrow $d$-bands were 
recently observed by ARPES \cite{mitr}. A simple tight-binding model explaining 
the physical properties at the MI transition includes a broad $A_{1g}$ band 
arising from the direct overlap of $3d_{z^2}$ orbitals along the $c$-axis, and a 
doubly degenerate $E_{g}$ narrow band created by the small V(3$d$)-S(3$p$) 
hybridization \cite{mas,bar}. The width and relative occupancy of the bands 
remain to be determined experimentally. An X-ray diffraction study \cite{fag} 
suggests a nearly equal filling of the $A_{1g}$ and $E_{g}$ bands. A tuning 
mechanism for the occupancy of these two bands was recently proposed \cite{lec}.

In this letter, we study the effects of chemical pressure on the magnetic, 
transport and structural properties of the system. At a critical value, 
$x$=0.07, an abrupt onset of FM order is found concomitant to a disappearance of 
the AF-like cusp at the MI transition and to a collapse of the unit cell at room 
temperature. A progressive reduction of $T_{MI}$ and smearing of the AF-like 
cusp with increasing $x$ is also observed. The observation of a quantum phase 
transition (QPT) from a non-FM to a FM low-temperature state shows that the 
incipient FM order in the parent compound is stabilized by chemical pressure. 
The second result suggests a coexistence of FM and metallic phases at high 
pressure \cite{bar}. We discuss the possibility that the two phases coincide in 
the absence of disorder. We argue that the collapse of the unit cell is due 
to a sudden redistribution of electron charge between the relevant $A_{1g}$ and 
$E_{g1}$ bands. 

A series of \bsvs~ single crystals with $x$ = 0.032, 0.053, 0.064, 0.068, 0.097, 
0.126, and 0.183 were grown as described elsewhere using a solid state method 
\cite{gau}. Attempts to grow crystals with larger $x$ values were unsuccessful. 
At $x$ = 0.183, diffraction data indicate a significant amount of disorder and 
stacking faults, thus $x \approx$ 0.18 appears to be close to the solubility 
limit. The needle-shaped crystals, 0.1 - 0.3 mm long, are aligned along the $c$-
axis. The room temperature crystal structure, the Sr/Ba composition and the 
sulfur content were determined using X-ray diffraction. Within the experimental 
resolution, the crystal symmetry remains $P6_3/mmc$ for all $x$ values, as in the 
unsubstituted compound. The structural refinement indicates no sulfur deficiency 
for any $x$ with a statistical uncertainty $\delta \pm$ 0.05 or better. As a 
further check of the sulfur stoichiometry, a post-sulfurisation treatment in 
evacuated quartz ampoules at 450$^\circ$C for 3 days is found to affect neither 
the structural nor the physical properties. Thus, the results reported are 
exclusively ascribed to the Sr-induced chemical pressure.

The crystals were investigated by means of dc electrical resistivity under 
pressure, $\varrho(T,P)$, up to 2 GPa, and dc magnetization, $M(T)$. $\varrho$ 
was measured in a four contact bar configuration in self-clamping pressure cells 
filled with kerosene as pressurizing agent. Since the needle-shaped crystals 
grow along the $c$-axis, the $c$-axis resistivity, $\varrho_c$, was measured. 
$M$ was measured in the zero-field and field-cooling (ZFC, FC) modes at 100 
oersted using a commercial RF SQUID magnetometer. Because of the small crystal 
size, the sample weight could not be determined. Yet, the sample sizes are 
comparable and the relative magnitude of $M$ is significant.

A set of $\varrho_c(T,P)$ curves is shown in Fig. 1 for $x$=0.126. 
Similarly to the case of the unsubstituted compound \cite{for} ($x$=0), a 
progressive reduction of $T_{MI}$ with increasing pressure is found. $T_{MI}$ 
was estimated as the peak of the $d$ln$\varrho/d(1/T)$ vs. $T$ plot \cite{for}. 
The effect of $T_{MI}$ reduction with increasing $x$ and with increasing 
pressure is apparent in the $P-T$ diagram of Fig. 2 that summarizes the 
$\varrho$ data. For $x \approx$ 0.12, the insulating state is suppressed at 
$\approx$ 1 GPa, i.e. half of the value for the unsubstituted compound \cite{for}. 
By comparing the $\varrho_c(T,P)$ curves in the inset of Fig. 1, one notes that 
the $T_{MI}$ reduction is accompanied by an increase of the slope, $d\varrho_c/dT$, 
with increasing pressure. This trend is common to all samples and indicates a 
pressure-induced enhancement of the metallic properties, although the system remains 
a bad metal. By comparing the $\varrho(T)$ curves for different $x$, we observed 
that the residual resistivity increases with increasing $x$ \cite{bar2}. This trend 
indicates that the Sr-substitution creates a substantial amount of disorder, 
in agreement with diffraction data.
\begin{figure}
\includegraphics[width=8cm]{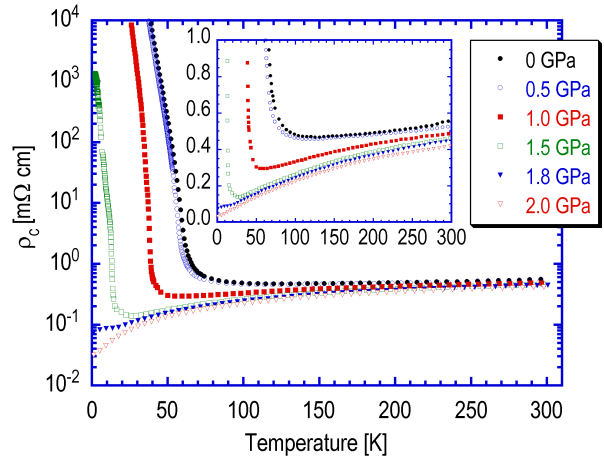}
\caption{$c$-axis resistivity curves as a function of temperature and pressure 
for $x$=0.126. One notes the steady reduction of the metal-insulator transition 
temperature $T_{MI}$, as pressure increases.}
\label{rhovsTP}
\end{figure}
\begin{figure}
\includegraphics[width=8cm]{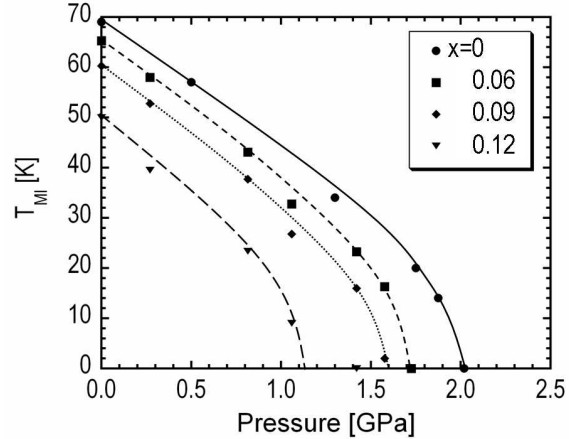}
\caption{Progressive reduction of $T_{MI}$ upon $x$ from the full set of the 
$\varrho_c(T,P)$ data. The $x=0$ data are taken from ref. \cite{for}. Lines are 
a guide to the eye.}
\label{T_MIvsx}
\end{figure}

The $T_{MI}$ reduction is correlated with the Sr--induced structural changes of 
the $a$- and $c$-axis as shown in Fig. 3. One notes that the smaller Sr ion 
induces a shrinking of the unit cell, as expected. Two features are though 
unusual: 1) at low substitution levels, $x \lesssim$ 0.05, the shrinking is 
negligible; 2) at higher $x$ values, the shrinking is large and occurs suddenly, 
first along the $c$-axis at $x \approx$ 0.07 and then in the $ab$-plane at $x 
\approx$ 0.10. The latter shrinking is largest and corresponds to a collapse of 
the chains of VS$_6$ octahedra one against the other. The $c$-axis increase at 
$x \gtrsim 0.10$ is consistent with the previous observation of disorder and 
stacking faults in this $x$ range. In spite of this, the cell volume continues 
to decrease with $x$, for the shrinking of the $a$-axis dominates. The sudden 
drop of the $a$- and $c$-axis at $x \approx$ 0.07 appears to be a room 
temperature signature of a phase transition at low temperature, as discussed below.
\begin{figure}
\includegraphics[width=8cm]{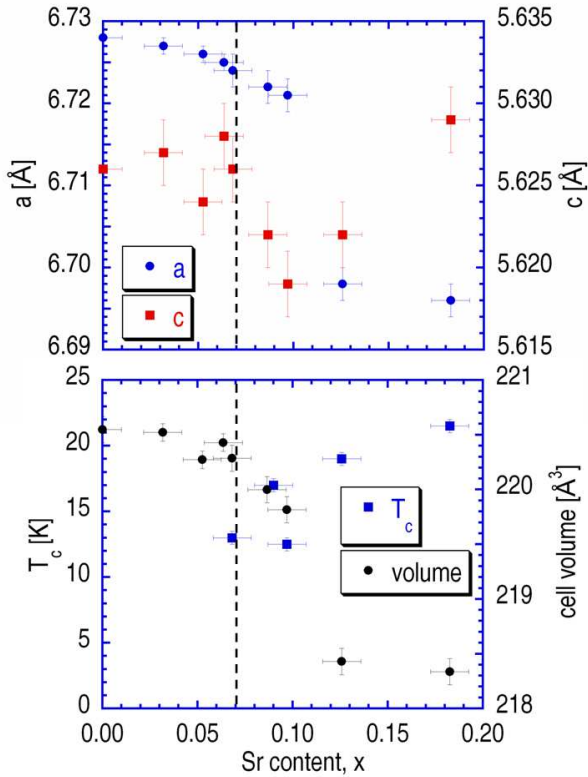}
\caption{Top panel: $a$- and $c$-axis variations of the hexagonal unit cell of 
\bsvs~ with $x$. Note the drop of the $c$-axis at $x \approx$ 0.07 and the 
progressive reduction of the $a$-axis followed by a large drop at $x \approx$ 
0.10. Error bars are standard deviations of the data refinement. Bottom panel: 
correlation between cell volume reduction and Curie temperature, $T_C$, and $x$. 
The vertical line at $x$ = 0.07 marks the onset of the cell volume drop and of 
the FM order.}
\label{cellvsx}
\end{figure}

By analyzing the structural data, the V-S bond distance in the VS$_6$ octahedra 
is found to be the only parameter that scales with $T_{MI}$. A linear regression 
of the $T_{MI}$ vs. V-S data enables to predict that the M-I transition would be 
suppressed for V-S=2.372 \AA. This result is in agreement with the observation 
of $T_{MI}$ reduction upon hydrostatic pressure \cite{for}. One concludes that 
the volume reduction induced by either chemical or hydrostatic pressure 
stabilizes the metallic state. The $T_{MI}$ reduction observed in our \bsvs~ 
samples is correlated to the magnetization data as follows. In Fig. 4 we report 
a selection of $M(T)$ curves for three different $x$ values in the $x$=0.06-0.10 
region, that turns out to be critical, as shown below. For $x \lesssim 0.07$, no 
significant difference between ZFC and FC curves is observed, so only the former 
case is considered. The AF-like cusp of $\chi$ characteristic of the unsubstituted 
compound \cite{tak,mas} remains visible at $T_{MI}$, although smeared by disorder. 
By taking the cusp as the magnetic signature of $T_{MI}$, the $T_{MI}$ vs. $x$ 
magnetization data are found to coincide with the resistivity ones discussed 
before. The two data sets are reported in Fig. 5 and show the same monotonic 
decrease of $T_{MI}$ with $x$. Also the upturn of the $M(T)$ curve at $T_X$ 
characteristic of the unsubstituted compound \cite{mih} remains visible for $x 
\lesssim 0.07$. Contrary to the case of the AF-like cusp, the temperature of the 
upturn hardly varies with $x$ but also the upturn becomes smeared with $x$. At 
$x_{cr}=0.07$, both the cusp and the upturn disappear and a FM response suddenly 
appears with $T_C$=12 K, determined as the maximum of the derivative of the 
$M(T)$ curve.
\begin{figure}
\includegraphics[width=8cm]{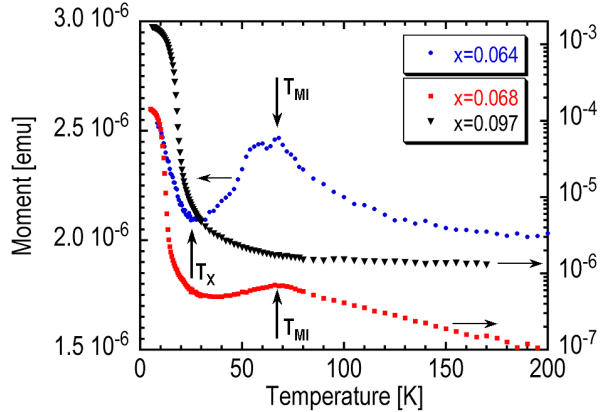}
\caption{ZFC magnetization for three different $x$-values showing the 
disappearance of the AF-like cusp at $T_{MI}$ and of the upturn at $T_X$ 
concomitant to the onset of FM order at $x \approx 0.07$. Note the different 
vertical scales.}
\label{MvsT}
\end{figure}

The above observations lead to conclude that the MI transition is not 
necessarily associated with the magnetic correlations. This contrasts to the 
transition at $T_x$ \cite{bar,bar2}, characterized by a FM order along the chains, 
in agreement with neutron diffraction data \cite{nak} and with the FM above 
$x_{cr}$ presented in this letter. The analysis of several crystals with $x 
\approx 0.07$ confirms the sudden character of the onset of the FM order and its 
coincidence with the disappearance of the AF-like cusp of $\chi$ and an equally 
sudden drop of cell volume. In the bottom panel of Fig. 3, we show the correlation 
between volume reduction and onset of FM order upon $x$. At $x_{cr} > 0.07$, $T_C$ 
slowly increases and equals 22 K at $x \approx$ 0.18.

Fig. 5 summarizes all data discussed above and serves as electronic phase 
diagram. The AF-like borderline is derived from the estimation of $T_X$ as 
upturn of the magnetization. The upturn being increasingly smeared with $x$, as 
it is apparent from Fig. 4, we believe that a long range order in the ab-plane 
no longer exists in the substituted compounds near $x_{cr}$. Thus, the sudden 
stabilization of the FM order at $x_{cr}$ and at low temperature is a quantum 
phase transition that distinguishes the FM order from a phase with AF-like 
correlations in the ab-plane that become weaker with increasing $x$. The drop 
of cell volume at $x_{cr}$ appears as a structural signature of the electronic 
instability at room temperature.
\begin{figure}
\includegraphics[width=8cm]{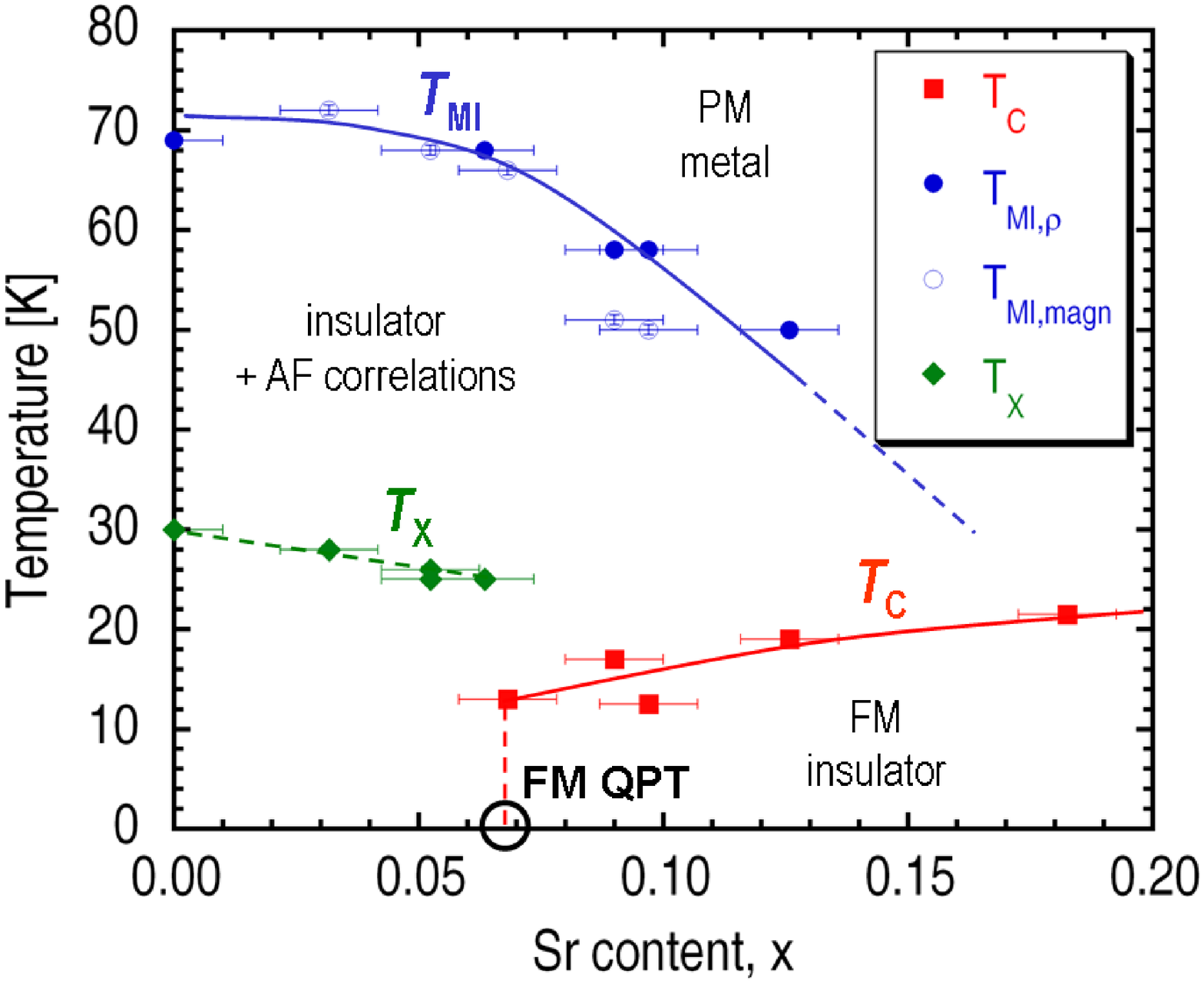}
\caption{Electronic phase diagram of Ba$_{1-x}$Sr$_x$VS$_3$ at ambient pressure. 
Full and open circles indicate the $T_{MI}$ values determined from, 
respectively, resistivity and magnetization data. The smearing of the AF-like 
cusp at $T_X$ upon $x$ discussed in the text is indicated as a broken line. The 
vertical line at $x$=0.07 marks the abrupt onset of ferromagnetic (FM) order. 
Lines are a guide to the eye.}
\label{phasediagram}
\end{figure}

In conclusion, we studied the effects of chemical pressure on the MI and AF-FM 
competitions in the BaVS$_3$ system on a series of \bsvs~ single crystals. The 
data of resistivity under pressure show that the Sr-induced chemical pressure 
stabilizes the metallic phase. Consequently, by increasing $x$ from 0 to 0.12 
the $T_{MI}$ is reduced from 69K down to 50 and the critical pressure for 
the suppression of the MI transition at 0K decreases from 2 to 1 GPa. At 
$x_{cr}$ = 0.07, we found a sudden shrinking of the unit cell concomitant to the 
disappearance of the AF-like susceptibility cusp at $T_{MI}$ and to the 
appearance of a full FM order at low temperatures. This transition confirms the 
small energy difference between the AF-like and FM orderings of the V$^{4+}$ 
ions calculated \emph{ab initio} \cite{san}. By extrapolating the $T_{MI}$ vs. 
V-S bond data, a metallic state at 0 K is predicted for a distance of 2.372 \AA.
The present data suggest that, at higher $x$ or pressure, the FM and metallic 
state should coexist \cite{bar}, as in BaVSe$_3$ \cite{yam,bar2}. The question 
arises as to the mechanism leading to the FM order in our case. The absence of 
sulphur deficiency rules out the Hund mechanism proposed for sulphur-deficient 
compounds \cite{yam2}. The Sr/Ba chemical disorder would favor a double-exchange 
mechanism, thus one should study samples without disorder. As disorder tends to 
stabilize the insulating state, the reduction of $T_{MI}$ reported here is 
expected to be even stronger in the absence of disorder. The MI quantum phase 
transition predicted at $x \approx 0.15$ from the data of Fig. 5 would then 
approach the FM QPT at $x=0.07$. In the extreme case, the two points may merge, 
which would be important from a fundamental point of view.

\begin{acknowledgments}
The authors thank M. Marezio for useful discussions and T. Besagni for technical 
assistance and acknowledge financial support provided by the \textit{Consiglio 
Nazionale delle Ricerche}.
\end{acknowledgments}

\end{document}